\documentclass[12pt]{article}
\usepackage{amsmath,amssymb }
\usepackage{fancybox}
\usepackage{graphicx,psfrag,epsf}
\usepackage{enumerate}
\usepackage{graphicx,psfrag}
\usepackage{multirow}
\usepackage{epsfig}
\usepackage{subfigure}
\usepackage{theorem}
\usepackage{natbib,psfrag}
\usepackage[usenames,dvipsnames]{color}
\usepackage{todonotes}
\usepackage{xcolor}
\usepackage{algorithm2e}
\usepackage{tabularx}
\usepackage{mathtools}
\newcommand{\pdf}{0}
\if1\pdf
\usepackage[pdftex]{graphicx}
\else
\usepackage{graphicx}
\fi

\newcommand{\blind}{0}

\def\mb#1{\setbox0=\hbox{$#1$}
  \kern-.025em\copy0\kern-\wd0
  \kern.05em\copy0\kern-\wd0
  \kern-.025em\raise.0em\box0}

\makeatletter
\newcommand*\dashline{\rotatebox[origin=c]{90}{$\dabar@\dabar@\dabar@$}}
\makeatother

\newtheorem{assumption}{Assumption}
\newtheorem{theorem}{Theorem}
\newtheorem{lemma}{Lemma}

\newtheorem{definition}{Definition}

\DeclareMathOperator*{\argmax}{arg\,max}
\DeclareMathOperator*{\argmin}{arg\,min}

\newcommand{\RN}[1]{%
	\textup{\uppercase\expandafter{\romannumeral#1}}%
}

\def\mb#1{\setbox0=\hbox{$#1$}
	\kern-.025em\copy0\kern-\wd0
	\kern.05em\copy0\kern-\wd0
	\kern-.025em\raise.0em\box0}
\def\0{{\bf 0}}


\begin{document}

\def\spacingset#1{\renewcommand{\baselinestretch}%
{#1}\small\normalsize} \spacingset{1}


\if0\blind
{
  \title{\bf Classification of Categorical Time Series Using the Spectral Envelope and Optimal Scalings}
  \author{Zeda Li*\\
    Baruch College, The City University of New York\\
     \\
    Scott A. Bruce\footnote{Both authors contributed equally to this work.}\\
    Department of Statistics, George Mason University\\
    \\
    Tian Cai\\
    Graduate Center, The City University of New York}
      \date{}
  \maketitle
} \fi

\if1\blind
{
  \bigskip
  \bigskip
  \bigskip
  \begin{center}
    {\LARGE\bf Title}
\end{center}
  \medskip
} \fi

\bigskip
\begin{abstract}
This article introduces a novel approach to the classification of categorical time series under the supervised learning paradigm. To construct meaningful features for categorical time series classification, we consider two relevant quantities: the spectral envelope and its corresponding set of optimal scalings. These quantities characterize oscillatory patterns in a categorical time series as the largest possible power at each frequency, or {\it spectral envelope}, obtained by assigning numerical values, or {\it scalings}, to categories that optimally emphasize oscillations at each frequency. Our procedure combines these two quantities to produce an interpretable and parsimonious feature-based classifier that can be used to accurately determine group membership for categorical time series.
Classification consistency of the proposed method is investigated, and simulation studies are used to demonstrate accuracy in classifying categorical time series with various underlying group structures.  Finally, we use the proposed method to explore key differences in oscillatory patterns of sleep stage time series for patients with different sleep disorders and accurately classify patients accordingly.  
\end{abstract}

\noindent%
\vfill

\newpage
\spacingset{1.5} 
\section{Introduction}

\begin{figure}[ht]
		\centering
		\includegraphics[scale=.6]{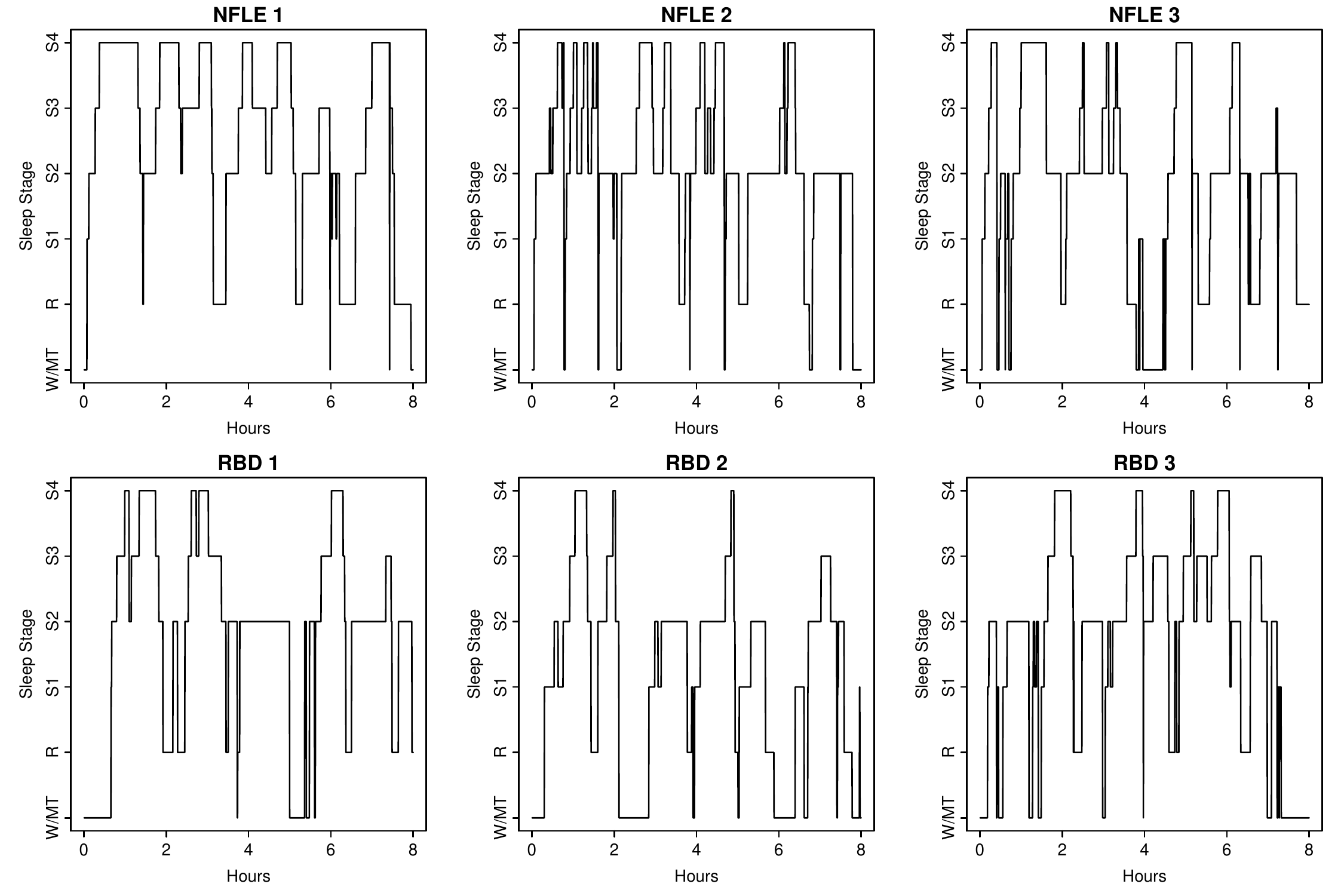}
		\caption{\small Sleep stage time series from six sleep study participants: three NFLE patients (top row) and three RBD patients (bottom row).}
				\label{sleepstages}
\end{figure}

Categorical time series are frequently observed in a variety of fields, including sleep medicine, genetic engineering, rehabilitation science, and sports analytics \citep{stoffer2000}. In many applications, multiple realizations of categorical time series from different underlying groups are collected in order to construct a classifier that can accurately identify group membership.
As a motivating example, we consider a sleep study in which participants with different types of sleep disorders are monitored during a night of sleep via polysomnography in order to understand important clinical and behavioral differences among these sleep disorders.  During sleep, the body cycles through different sleep stages: movement/wakefulness, rapid eye movement (REM) sleep, and non-rapid eye movement (NREM) sleep, which is further divided into light sleep (S1,S2) and deep sleep (S3, S4).  Our analysis focuses on two particular sleep disorders, nocturnal frontal lobe epilepsy (NFLE) and REM behavior disorder (RBD), for which differential diagnosis is especially challenging due to a significant overlap in their associated clinical and behavioral characteristics \citep{tinuper2017nocturnal}.  For example, NFLE and RBD patients both exhibit complex, bizarre motor behavior and vocalizations during sleep.  However, we posit that differences in sleep cycling behavior may still exist due to fundamental differences in the sleep disruption mechanisms of NFLE and RBD.  The goal of our analysis is to investigate potential differences in sleep cycling behavior for NFLE and RBD patients and use this information to accurately classify patients accordingly.  This data-driven classification can potentially improve accuracy in differential diagnoses of NFLE and RBD in patients presenting clinical and behavioral characteristics common to both conditions.
Figure~\ref{sleepstages} displays examples of study participants' full night sleep stages series from two different groups. 

In the statistical literature, classification methods for multiple real-valued time series have been well-studied; see \cite{shumway2016} for a review.  However, classification of categorical time series has not received much attention.  The majority of statistical methods for categorical time series analysis have been developed for analyzing a single categorical time series.  Some examples include the Markov chain model of \cite{billingsley1961},  the link function approach of \cite{fahrmeir1987}, the likelihood-based method of \cite{fokianos1998}, and the spectral envelope approach for analyzing a single time series introduced in \cite{stoffer1993}.  A comprehensive discussion of this research direction can be found in \cite{fokianos2003}. 
More recently, \cite{krafty2012} introduced the spectral envelope surface for quantifying the association between the oscillatory patterns of a collection of categorical time series and continuous covariates. However, it is not immediately useful for classification. To the best of our knowledge, this article presents the first statistical approach for supervised classification of multiple categorical time series. 

In the computer science literature, however, many methods have been developed to classify string-valued time series, which can also be used for classification of categorical time series. These include the minimum edit distance classifier with sequence alignment \citep{gonzalo2001,jurafsky2009}, Markov chain-based classifiers \citep{mukund2002}, the Haar Wavelet classifier \citep{Aggarwal2002}, and the state-of-the-art sequence learner that uses a gradient-bounded coordinate-descent algorithm for efficiently selecting discriminative subsequences and then uses logistic regression for classification \citep{ifrim2011}. These methods are black-box in nature and offer little help in understanding key differences among groups. On the other hand, the proposed method addresses the classification problem using the spectral envelope and optimal scalings, which provide low-dimensional, interpretable summary measures of oscillatory patterns and traversals through categories.  These patterns are often associated with scientific mechanisms that distinguish different groups and also produce lower classification error compared to state-of-the-art computer science methods like sequence learner.




Many classifiers for real-valued time series rely on feature extraction, a process in which low-dimensional summary quantities are constructed that capture essential features of the underlying groups.  These quantities are then used to develop feature-based distance measures, such as the Kullback-Leibler distance and squared quadratic distance, which can be used to 
measure differences between groups and time series of unknown group membership.  Training data can then be used to estimate group-level quantities and construct a classifier that minimizes the distance between time series and their predicted group \citep{huang2004, shumway2016}.
This type of approach cannot be easily extended to the classification of categorical time series due to the difficulty in obtaining low-dimensional features. To this end, we propose using the spectral envelope and its corresponding set of optimal scalings \citep{stoffer1993} as low-dimensional, interpretable features for differentiating groups of categorical time series. 
Use of these features is motivated by noticing that most categorical time series can be represented in terms of their prominent oscillatory patterns, characterized by the spectral envelope, and by the set of mappings from categories to numeric values that accentuate specific oscillatory patterns, characterized by the optimal scalings.

\begin{figure}[ht]
	\centering
	\includegraphics[height=3.6in]{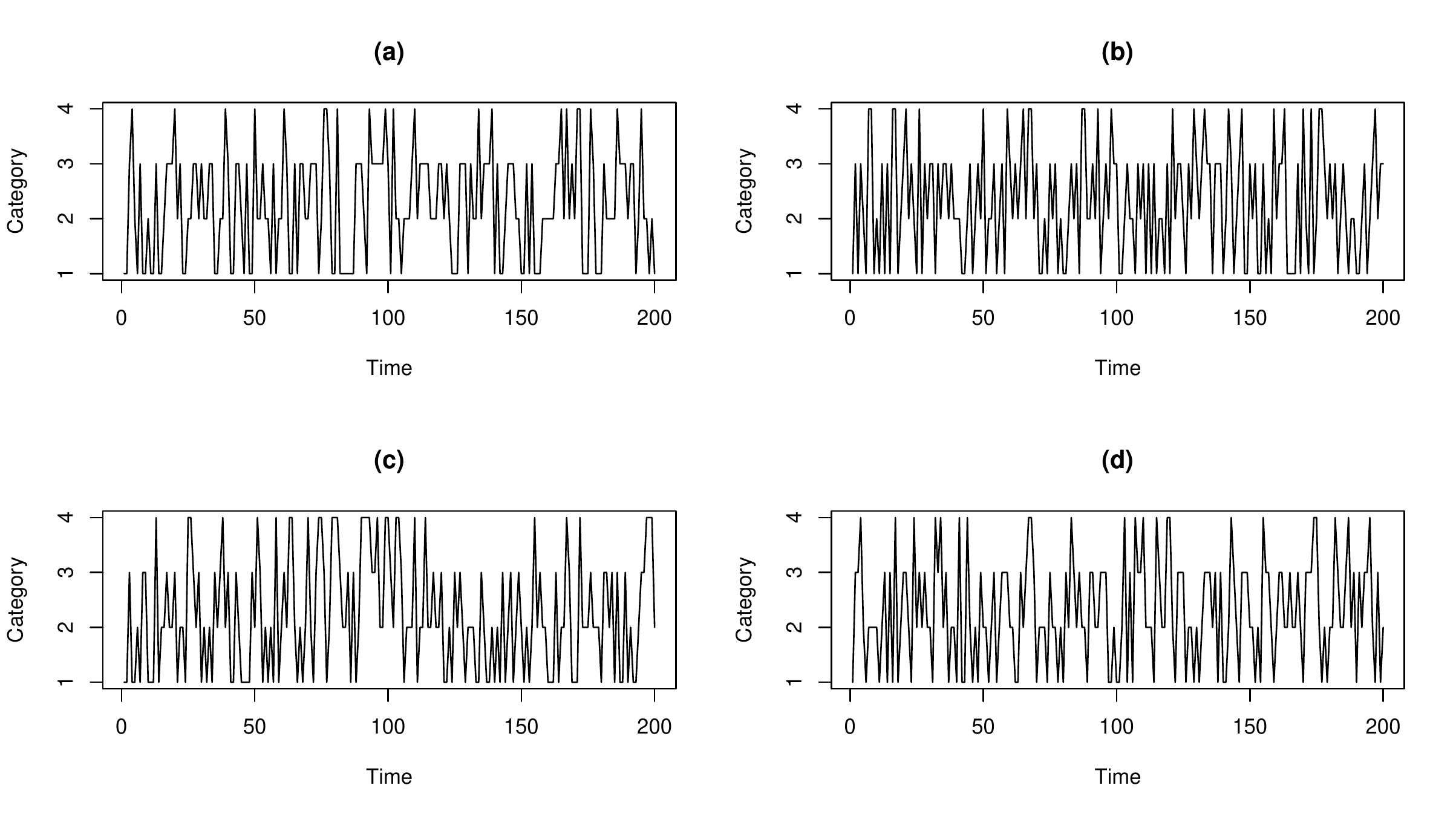}
	\caption{\small Four simulated categorical time series: (a) and (b) have the same dominating categories but different cyclical patterns; (c) and (d) have the same frequency patterns but different dominating categories.}
	\label{simulated}
\end{figure}

For example, Figures \ref{simulated}(a) and \ref{simulated}(b) display two categorical time series with similar traversals through categories, but different oscillatory patterns.  More specifically, the time series in Figure \ref{simulated}(b) cycles between categories \textit{faster} than the time series in Figure \ref{simulated}(a).  On the other hand, Figures \ref{simulated}(c) and \ref{simulated}(d) display two categorical time series with similar oscillatory patterns, but different traversals through categories.  More specifically, the time series in Figure \ref{simulated}(c) spends approximately \textit{equal} amounts of time in each category, while the time series in Figure \ref{simulated}(d) spends more time in categories 2 and 3.  Moreover, Figure \ref{compare} displays the estimated spectral envelope for the two series in Figures \ref{simulated}(a) and \ref{simulated}(b) and the optimal scalings for the two series in Figures \ref{simulated}(c) and \ref{simulated}(d).  The spectral envelope and optimal scalings clearly reflect the corresponding differences between these series.  In particular, the spectral envelope indicates more high frequency power for the time series in Figure \ref{simulated}(b) since it cycles between categories faster relative to the time series in Figure \ref{simulated}(a).  Also, the optimal scalings for the time series in Figure \ref{simulated}(c) and Figure \ref{simulated}(d) are quite different, reflecting the different traversals over categories resulting in different distributions of time spent in categories.

The proposed method is briefly described as follows. For each time series to be classified, we represent it as a vector-valued time series through the use of indicator variables. The smoothed spectral density matrix of this vector-valued time series is then obtained, and the spectral envelope and optimal scalings at each frequency are computed from the estimated spectral matrix. Then, the spectral envelope and optimal scalings for each group are estimated respectively via training data. 
The proposed feature, which is used to estimate the distance from each group,  is obtained by adaptively summing the differences in the spectral envelope and optimal scalings. Finally, time series with unknown group membership are assigned to groups with the most similar features (i.e. minimum distance). Under the proposed framework, we show that the misclassification probability is bounded as long as the spectral density matrix estimator is consistent. The procedure is demonstrated to perform well in simulation studies and a real data analysis. 

The remainder of the paper is organized as follows.  Section \ref{sec:env} provides definitions of the spectral envelope and optimal scalings and corresponding estimators.  Section \ref{sec:class} introduces the proposed classification procedure and its theoretical properties.  Section \ref{sec:sim} provides detailed simulation studies, which explore the empirical properties of the proposed method and compares with the state-of-the-art sequence learner classifier.  Section \ref{sec:app} details the application of the proposed classifier to the analysis of sleep stage time series to better understand and accurately classify sleep disorders. Section \ref{sec:end} provides some closing discussions and impactful extensions of this work.

\section{The Spectral Envelope and Optimal Scalings}
\label{sec:env}
\subsection{Definition}
\label{sec:defn}
Let $X_t$, for $t=0, 1, 2, \ldots,$ be a categorical time series with finite state-space $\mathcal{C}=\{c_1, c_2, \ldots, c_m\}$. We assume that $X_t$ is stationary 
such that $\{X_1,X_2,\ldots,X_{t}\} $ $\,{\buildrel d \over =}\, \{X_{1+h},X_{2+h},\ldots,X_{t+h}\}$ for $h \ge 0$ 
and $\inf_{\ell=1,2,\ldots,m} \mathrm{P}(X_t = c_{\ell})>0$ so that there are no absorbing states. In order to obtain a quantifiable measure of oscillatory patterns for categorical time series, a typical way is to consider a real-valued time series, $X_t(\beta)$, 
obtained by assigning numerical values, or scalings, to categories
such that $\beta = (\beta_1, \beta_2, \ldots, \beta_m)' \in \mathbb{R}^m$ and $X_t(\beta)=\beta_{\ell}$ when $X_t = c_{\ell}$. We assume that $X_t(\beta)$ has 
a continuous and bounded spectral density
$$
f_x(\omega;\beta) = \sum_{h=-\infty}^{\infty}\text{Cov}[X_t(\beta), X_{t+h}(\beta)] \exp(- 2 \pi i \omega h).
$$

Let $V_x(\beta)$ be the variance of the scaled time series $X_t(\beta)$, the spectral envelope is then defined as the maximal normalized spectral density, $f_x(\omega;\beta)/V_x(\beta)$, among all possible scalings not proportional to $1_m$  at frequency $\omega$, where $1_m$ is the $m$-dimensional vector of ones. Scalings that assign the same value to each category are excluded since $V_x(\beta)$ is zero and the normalized power spectrum is not well defined.  Formally, we define the spectral envelope and set of optimal scalings for frequency $\omega$ as
\begin{equation*}
\lambda(\omega) = \max_{\beta \in \mathbb{R}^m 	\setminus \{1\}} \frac{f_x(\omega;\beta)}{V_x(\beta)}, ~ B(\omega) =\argmax_{\beta \in \mathbb{R}^m 	\setminus \{1\}} \frac{f_x(\omega;\beta)}{V_x(\beta)},
\end{equation*}
respectively, where $\{1\}$ is the subspace of $\mathbb{R}^m$ that is proportional to $1_m$. The spectral envelope, $\lambda(\omega)$, is the largest proportion of the variance  that can be obtained at frequency $\omega$ for different possible scalings,  such that $f_{x}(\omega, \beta) \le \lambda(\omega)$ $\forall \beta \in \mathbb{R}^m 	\setminus \{1\}$. The spectral envelope characterizes important oscillatory patterns in categorical time series. 

\begin{figure}[ht]
	\centering
	\includegraphics[height=4.0in]{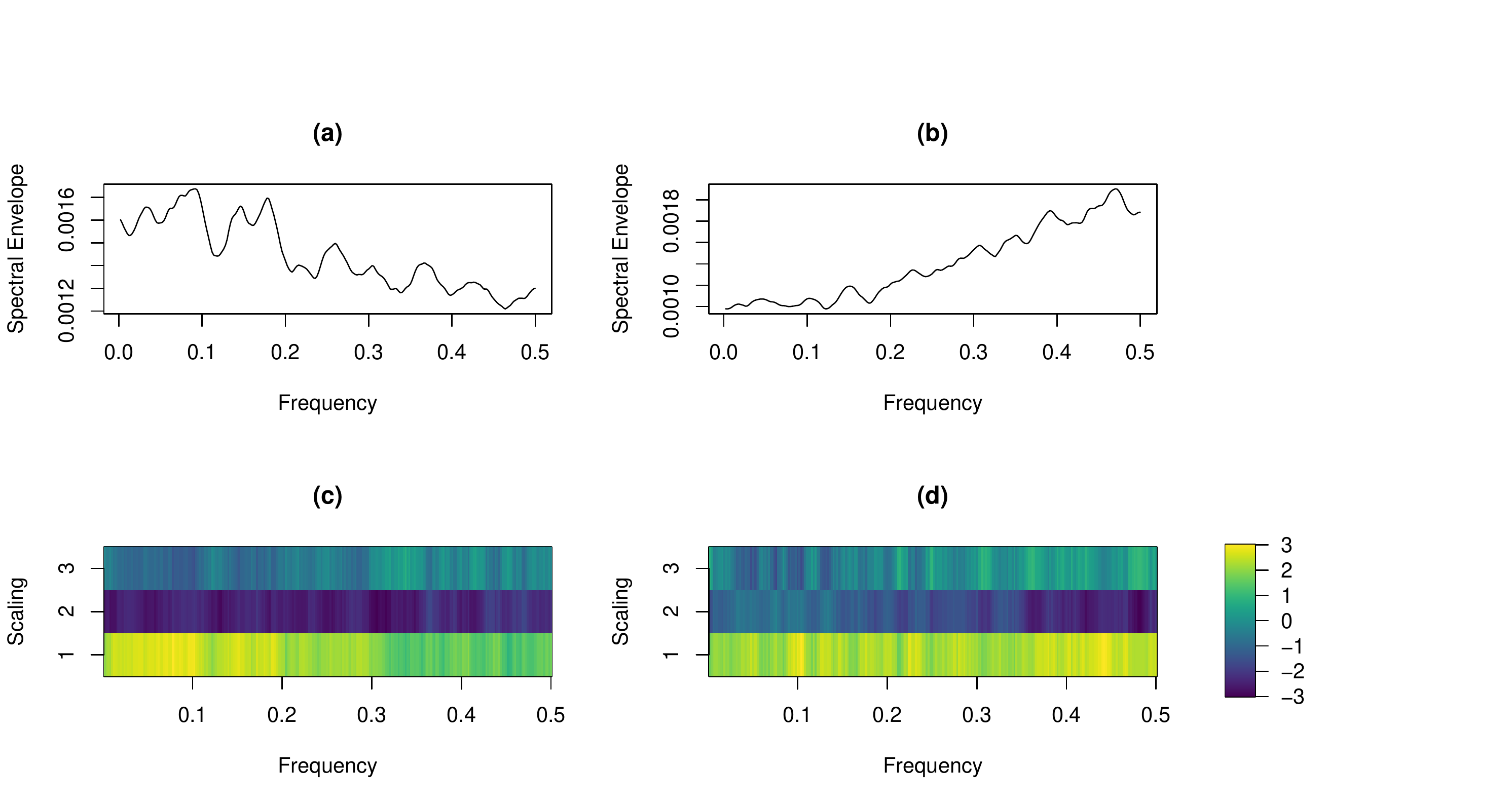}
	\caption{\small (a) and (b): The spectral envelopes of the time series shown in panels (a) and (b) of Figure 2; (c) and (d): The scalings of the time series presented in panels (c) and (d) of Figure 2. }
	\label{compare}
\end{figure}

For illustration, Figures~\ref{compare}(a) and \ref{compare}(b) display the estimated spectral envelopes for time series displayed in Figures~\ref{simulated}(a) and \ref{simulated}(b) respectively.
It can be seen that the time series in Figure~\ref{simulated}(a), which oscillates more slowly than the time series in Figure~\ref{simulated}(b), 
has more power in the estimated spectral envelope at lower frequencies. The set of optimal scalings that maximize the normalized spectral density at frequency $\omega$, $B(\omega)$, provides important information about the traversals through categories associated with prominent oscillatory patterns at frequency $\omega$.  
For further illustration, Figures~\ref{compare}(c) and \ref{compare}(d) display the estimated optimal scalings for time series displayed in Figures~\ref{simulated}(c) and \ref{simulated}(d) respectively. 
The optimal scalings in Figure~\ref{compare}(d) for categories 2 and 3 are similar at lower frequencies ($\omega<0.2$), but the optimal scalings in Figure~\ref{compare}(c) for categories 2 and 3 are different at lower frequencies.
This is because the corresponding time series in Figure~\ref{simulated}(d) visits categories 2 and 3 more frequently than the time series in Figure~\ref{simulated}(c).

\subsection{Computation Through Reparameterization} 
\label{sec:reparam}
A common approach to the analysis of any type of categorical data is to represent it in terms of random vectors of indicator variables. Similar to the formulations used in \cite{stoffer1993, krafty2012}, we define the $(m-1)$-dimensional stationary time series $Y_t$, which has a one in the $\ell$th element if $X_t = c_{\ell}$ for $\ell=1,\ldots, m-1$ and zero elsewhere. This representation is equivalent to setting the category $c_m$ as the reference category and restricting the set of optimal scalings 
to a lower-dimensional space.  The assumption that $f_x(\omega, \beta)$ is continuous is necessary and sufficient for ensuring that $Y_t$ has a continuous spectral density, which is defined as
\begin{equation*}
f_{y}(\omega) = \sum_{h = -\infty}^{\infty}\text{Cov}[Y_t, Y_{t+h}] \exp(- 2 \pi i \omega h).
\end{equation*}
The spectral density $f_{y}(\omega)$ is a positive definite Hermitian $(m-1) \times (m-1)$ matrix.
 We assume $f_{y}(\omega)$ and the variance of $Y_t$, $V_y=\text{Var}(Y_t)$, are non-singular for all $\omega \in \mathbb{R}$ \citep{brillinger2001}. Formally, we define the spectral envelope and the corresponding set of optimal scalings used in our proposed classification algorithm as follows.
\begin{definition}
	For $\omega \in \mathbb{R}$, the spectral envelope, $\lambda(\omega)$, is defined as the largest eigenvalue of 
	$$
	h(\omega) = V_y^{-1/2} f_y(\omega) V_y^{-1/2}.
	$$
	The $(m-1)$-variate vector of optimal scalings, $\gamma(\omega)$, is defined as the eigenvector associated with $\lambda(\omega)$. 
\end{definition}	

Several aspects of the definition should be noted. First, since the spectral density matrix is complex-valued and Hermitian with a skew symmetric imaginary component, for every $a \in \mathbb{R}^{m-1}$, we have $a' f_{y}(\omega) a = a' f_{y}^{re}(\omega) a$, where $f_{y}^{re}(\omega)$ is the real part of $f_{y}(\omega)$. Thus, the spectral envelope is equivalent to the largest eigenvalue of $h(\omega)^{re}= V^{-1/2} f_y^{re}(\omega) V^{-1/2}$. Second, a connection between the optimal scalings derived from this formulation and that defined in Section \ref{sec:defn} can be established \citep{krafty2012}. If $V_y^{1/2} \gamma(\omega)$ is an eigenvector of $h^{re}(\omega)$ associated with $\lambda(\omega)$, then  
\begin{eqnarray*}
	\begin{bmatrix}
		\gamma(\omega)\\
		0
	\end{bmatrix}	=
	\argmax_{\beta \in \mathbb{R}^m 	\setminus \{1\}} \frac{f_x(\omega;\beta)}{V_x(\beta)}.
\end{eqnarray*}
When the multiplicity of $\lambda(\omega)$ as an eigenvalue of $h^{re}(\omega)$ is one, there exists a unique $\gamma(\omega)$ such that $V^{1/2}_y \gamma(\omega)$ is an eigenvector of $\gamma(\omega)$ associated with $\lambda(\omega)$ where $\gamma(\omega)' V_y     \gamma(\omega) = 1$  and with the first nonzero entry of $V^{1/2} \gamma(\omega)$ to be positive.
Third, if there is a significant frequency component near $\omega$, then $\lambda(\omega)$ will be large, and the values of $\gamma(\omega)$ are dependent on the particular cyclical traversal of the series through categories that produces the value of $\lambda(\omega)$ at frequency $\omega$.

\subsection{Estimation}
\label{sec:est}
Consider a realization of a categorical time series, $X_{t}, t \ldots, T$, and its corresponding multivariate process realization $Y_{t}, t \ldots, T$ defined in Section \ref{sec:reparam}. Let $\hat{f}_y(\omega)$ represent the estimate of the spectral matrix $f_y(\omega)$. To allow for asymptotic development, we assume $Y_t$  is strictly stationary and that all cumulant spectra, of all orders, exist \citep[Assumption 2.6.1]{brillinger2001}. There is an extensive literature on estimation of the power spectral matrix. We use periodograms, or sample analogues of the spectrum
\begin{equation*}
I(s) = T^{-1} \left | \sum_{t=1}^{T} Y_t \exp(-2 \pi i s t/T) \right |^2, ~~s=1, \ldots T.
\end{equation*}
It is well known that the periodogram is an asymptotically unbiased but inconsistent estimator of the true spectral matrix. A common way to obtain a consistent estimator of the spectral matrix is to smooth periodogram ordinates over frequencies using kernels \citep{brillinger2001}. In this paper, we consider the smoothed periodogram estimator
\begin{equation*}
\hat{f}_{y}(\omega_s) =  \sum_{j=-B_T}^{B_T} W_{B_T, j} I(s + j),
\end{equation*}
where $\omega_s=s/T$ for  $s=1, \ldots, K=\lfloor(T-1)/2\rfloor$ are the Fourier frequencies,
 $2B_T+1$ is the smoothing span, and $W_{B_T, j}$ are nonnegative weights that satisfy the following conditions:
\begin{equation*}
W_{B_T, j} = W_{B_T, -j}, ~~ \sum_{j=-B_T}^{B_T} W_{B_T, j} = 1.
\end{equation*} 
Generally, the weights are chosen such that $W_{B_T, 0}$ is a decreasing function of $B_T$. It is known that $\hat{f}_y(\omega_k)$ is consistent if $B_T \rightarrow \infty$ and $B_T T^{-1} \rightarrow 0$ as $T \rightarrow \infty$ \citep{brillinger2001}. 
Given the sample spectral matrix $\hat{f}_y(\omega)$ and sample variance $\widehat{V}_y$, the estimate of the spectral envelope $\hat{\lambda}(\omega)$ is the largest eigenvalue of 
$
\hat{h}(\omega)^{re}= \widehat{V}^{-1/2} \hat{f}_y^{re}(\omega) \widehat{V}^{-1/2},
$
and the optimal scaling, $\hat{\gamma}(\omega)$, is the eigenvector of $\hat{h}(\omega)^{re}$ associated with $\hat{\lambda}(\omega)$. It should be noted that other approaches for nonparametric estimation of the spectral matrix, such as those in \cite{dai2004}, \cite{rosen2007}, and \cite{krafty2013}, can also be used. We use the kernel smoothing approach for computational efficiency and ease of theoretical exposition.

\section{The Classification Methods}
\label{sec:class}
Consider a population of categorical time series composed of $J \ge 2$ groups, $\Pi_1, \ldots \Pi_J$. Denote the $j$th group-level spectral envelope and $(m-1)$-variate scaling as $\Lambda^{(j)}(\omega)$ and $\Gamma^{(j)}(\omega)$ for $j=1,\ldots, J$ respectively. Suppose we observe $N=\sum_{j=1}^{J} N_j$ independent training time series of length $T$
and $R$ independent testing time series of length $T$, $X_r = \{ X_{r1}, \ldots, X_{rT} \}$, $r=1, \ldots, R$, with unknown group membership.
In this section, we introduce an adaptive algorithm for consistent classification.


\subsection{Classification via the Spectral Envelope}
\label{sec:class_se}

As shown in Figures~\ref{simulated} and \ref{compare}, groups of categorical time series may exhibit distinct oscillatory patterns. In this case, the spectral envelope, which characterizes dominant oscillatory patterns, can be used as a signature for each group and an important feature for categorical time series classification. We outline a classification procedure based on the spectral envelope below.
\begin{enumerate}
\item For each testing time series, compute the sample spectral envelope, $\hat{\lambda}^{(r)}(\omega_s)$, for $r=1, \ldots, R$, where $\omega_s =s/T$ are the Fourier frequencies with $s=1,\ldots, K$ and $K=\lfloor(T_{\ell}-1)/2\rfloor$. Denote $\hat{\lambda}^{(r)}= \{\hat{\lambda}^{(r)}(\omega_1), \ldots, \hat{\lambda}^{(r)}(\omega_K)\}'$ as a $K$-dimensional vector.

\item 	Compute 
\begin{equation}\label{dist1}
			D_{j,ENV}^{(r)} = || \hat{\lambda}^{(r)}-\Lambda^{(j)}||^2_2,\;
\end{equation}
where $||\cdot||_2$ is the $L_2$ norm.

\item Classify time series $X_{r}$ to the group $\Pi_j$ with the most similar spectral envelope such that $$\hat{g}_r = \argmin_j D^{(r)}_{j,ENV} \quad j=1,\ldots,J.$$
\end{enumerate}

Classification consistency can be established under the following assumptions. To aid the presentation, we consider the case of $J=2$ groups, $\Pi_1$ and $\Pi_2$, while similar results can be derived for $J>2$.
\begin{assumption}
	Each element of the $(m-1) \times (m -1)$ spectral density matrix $f_y(\omega)$ has bounded and continuous first derivatives. 
\end{assumption}	
\begin{assumption}
	$|| \Lambda^{(1)}-\Lambda^{(2)}||^2_2 \ge CT$ for a positive constant $C$.
\end{assumption}
Under Assumption 1, asymptotic consistency of the estimates $\hat{\lambda}(\omega)$ and $\hat{\gamma}(\omega)$ discussed in Section \ref{sec:est} can be established, and the largest eigenvalue of the spectral density matrix is continuous and bounded from above. Assumption 2 implies that the spectral envelopes of the two groups are well separated. The following theorem states the classification consistency of using the spectral envelope as a classifier.

\begin{theorem}\label{consist1}
	Under Assumptions 1-2, the probability of misclassifying $X_{r}$, a testing time series from group $\Pi_1$, to group $\Pi_2$, can be bounded as follows:
	$$
	P(D^{(r)}_{1, ENV} > D^{(r)}_{2, ENV}) = O(B_T^{2}T^{-2}),
	$$
	where $D^{(r)}_{1, ENV}$ and $D^{(r)}_{2, ENV}$ are defined in \eqref{dist1}.
\end{theorem}		

\subsection{Classification via Optimal Scalings}
\label{sec:class_os}

While the spectral envelope adequately characterizes dominant oscillatory patterns, it doesn't account for traversals through categories responsible for such oscillatory patterns.  Differences among groups may also be due to different traversals through categories that produce particular oscillatory patterns, which are characterized by optimal scalings for each frequency component.  
Similarly, we present a categorical time series classifier using optimal scalings below. 
\begin{enumerate}
    \item Compute the $(m-1)$-dimensional sample scaling, $\hat{\gamma}^{(r)}(\omega_s)$, of the testing time series $X_{r}$ for $r=1, \ldots, R$, where $\omega_s =s/T$ are the Fourier frequencies with $s=1,\ldots, K$ and $K=\lfloor(T_{\ell}-1)/2\rfloor$. Denote $\hat{\gamma}^{(r)}= \{\hat{\gamma}^{(r)}(\omega_1)', \ldots \hat{\gamma}^{(r)}(\omega_K)'\}'$ as a $K \times (m-1)$ matrix.
    
    \item Compute 
			\begin{equation}\label{dist2}
			D_{j,SCA}^{(r)} = || \hat{\gamma}^{(r)}-\Gamma^{(j)}||_F^2\;
			\end{equation}
		where $||\cdot||_F$ is the Frobenius norm.
	\item Classify time series $X_{r}$ to the group $\Pi_j$ with the most similar set of optimal scalings such that $$\hat{g}_r = \argmin_j D^{(r)}_{j,SCA} \quad j=1,\ldots,J.$$

\end{enumerate}
In addition to Assumption 1, the following assumption is necessary to establish the classification consistency of the scaling classifier, which indicates that the optimal scalings are well separated.
\begin{assumption}
	For fixed $m$ categories, $|| \Gamma^{(1)}-\Gamma^{(2)}||^2_F \ge CT$ for a positive constant $C$.
\end{assumption}
Theorem 2 states the consistency of classification based on the scalings.
\begin{theorem}\label{consist2}
	Under Assumptions 1 and 3, the probability of misclassifying $X_{r}$, a testing time series from group $\Pi_1$, to group $\Pi_2$, can be bounded as follows:
	$$
	P(D^{(r)}_{1, SCA} > D^{(r)}_{2, SCA}) = O(B_T^{2}T^{-2}),
	$$
	where $D^{(r)}_{1, SCA}$ and $D^{(r)}_{2, SCA}$ are defined in \eqref{dist2}.
\end{theorem}		

\subsection{Proposed Adaptive Envelope and Scaling Classifier}

The envelope classifier (Section \ref{sec:class_se}) works well in situations where oscillatory patterns are different among groups, while the scaling classifier (Section \ref{sec:class_os}) is effective when traversals through categories are distinct among groups.  However, in practice, different groups are likely to exhibit different oscillatory patterns and traversals through categories to some extent.  
Thus, it is desirable to construct an adaptive classifier that can automatically identify the extent to which groups are different with respect to their oscillatory patterns, traversals through categories, or both, and optimally classify time series accordingly. 
To this end, we propose a general purpose, flexible classifier that adaptively weights differences in the spectral envelope and optimal scalings in order to determine the characteristics that best distinguish groups and provide accurate classification.
Specifically, we consider the following distance of the $r$th testing time series to the $j$th group
\begin{equation}\label{dist3}
D^{(r)}_{j,EnvSca} = \kappa \frac{||\hat{\lambda}^{(r)}-\Lambda^{(j)}||^2_2}{||\hat{\lambda}^{(r)}||^2_2} +  (1-\kappa) \frac{||\hat{\gamma}^{(r)}-\Gamma^{(j)}||_F^2}{||\hat{\gamma}^{(r)}||_F^2},
\end{equation}
for $j=1, \ldots, J$ and $r=1, \ldots, R$. Since the spectral envelope $\hat{\lambda}^{(r)}$ is a $K$-dimensional vector and the scaling $\hat{\gamma}^{(r)}$ is $(m-1) \times K$ matrix, we rescale these distances by their corresponding norms.  The unknown tuning parameter $\kappa$ controls the relative importance of the spectral envelope and optimal scalings in classifying time series. Our proposed adaptive classification algorithm is presented in Algorithm \ref{algo:envsca}. 
\spacingset{1}
\begin{algorithm}[ht!]
	\KwData{$R$ independent testing time series , $X_r = \{ X_{r1}, \ldots, X_{rT} \}$ for $r=1, \ldots, R$.}
	\KwResult{Estimated group assignment for each testing time series, $\{\hat{g}_1, \ldots, \hat{g}_R\}$, where $\hat{g}_r \in (1, \ldots, J)$ for $r=1, \ldots, R$.}
	Step 1: Use Leave-one-out cross validation to select tuning parameter $\kappa$.
	
	Step 2:
	
	\For{$r=1, \ldots R$}{
		
		Convert the testing time series $X_{r}$ with $m$ categories into a $(m-1)$-dimensional time series $Y_{r}$ defined in Section \ref{sec:reparam} and compute the $(m-1) \times (m-1)$ matrix $\hat{h}(\omega)$ in Definition 1\;
		
		\vspace{0.1cm}
		
		Compute the sample spectral envelope, $\hat{\lambda}^{(r)}(\omega_s)$, of the testing time series $X_{r}$, where $\omega_s =s/T$ are the Fourier frequencies with $s=1,\ldots, K$ and $K=\lfloor(T_{\ell}-1)/2\rfloor$. Denote $$\hat{\lambda}^{(r)}= \{\hat{\lambda}^{(r)}(\omega_1), \ldots \hat{\lambda}^{(r)}(\omega_K)\}'$$ as a $K$-dimensional vector\;
		
		Compute the $(m-1)$-dimensional sample optimal scalings, $\hat{\gamma}^{(r)}(\omega_s)$, of the testing time series $X_{r}$, where $\omega_s =s/T$ are the Fourier frequencies with $s=1,\ldots, K$ and $K=\lfloor(T_{\ell}-1)/2\rfloor$. Denote $$\hat{\gamma}^{(r)}= \{\hat{\gamma}^{(r)}(\omega_1)', \ldots \hat{\gamma}^{(r)}(\omega_K)'\}'$$ as a $K \times (m-1)$ matrix\;
		
		\vspace{0.1cm}
		\For{$j=1, \ldots J$}{		
		Compute 
		\begin{equation*}
D^{(r)}_{j,EnvSca} = \kappa \frac{||\hat{\lambda}^{(r)}-\Lambda^{(j)}||^2_2}{||\hat{\lambda}^{(r)}||^2_2} +  (1-\kappa) \frac{||\hat{\gamma}^{(r)}-\Gamma^{(j)}||_F^2}{||\hat{\gamma}^{(r)}||_F^2}.		\end{equation*}
	}
		\vspace{0.1cm}
		
		Classify the time series $X_{r}$ to group $\Pi_j$ if $D^{(r)}_{j,EnvSca}$ is the smallest among all $D^{(r)}_{j,EnvSca}$ for $j=1, \ldots, J$, that is, $$\hat{g}_r = \argmin_j D^{(r)}_{j,EnvSca}.$$
		
	}
	\Return{$\{\hat{g}_1, \ldots, \hat{g}_R\}$}\;
	\caption{{\sc Envelope and Scaling Classifier} (EnvSca)}
	\label{algo:envsca}    
\end{algorithm}
\spacingset{1.5} 

Several remarks on the algorithm should be noted. First, the group-level spectral envelopes $\Lambda^{(j)}$ and optimal scalings $\Gamma^{(j)}$ are unknown in practice. We obtain $\Lambda^{(j)}$ and $\Gamma^{(j)}$ by averaging the sample spectral envelopes and sample optimal scalings across training time series replicates within the $j$th group, respectively. In particular, 
we replace $\Lambda^{(j)}$ and $\Gamma^{(j)}$ by their sample estimates
\begin{eqnarray*}
\hat{\Lambda}^{(j)}= \frac{1}{N_j} \sum_{k=1}^{N_j} \hat{\lambda}^{(j,k)},~~
\hat{\Gamma}^{(j)}= \frac{1}{N_j} \sum_{k=1}^{N_j} \hat{\gamma}^{(j,k)},
\end{eqnarray*}
for $j=1, \ldots, J$, where $\hat{\lambda}^{(j,k)}$ and $\hat{\gamma}^{(j,k)}$ are the estimated spectral envelope and optimal scalings of the $k$th training time series among group $j$, respectively. Second, we select the tuning parameter $\kappa$ by using a grid search through leave-one-out (LOO) cross-validation. In particular, let $\kappa \in (0, 0.1, 0.2, \ldots, 1)$.  The estimated $\hat{\kappa}$ corresponds to the value that produces the highest leave-one-out classification rate via Algorithm 1. Although a finer grid could be used as well, in our experience, using $\kappa \in (0, 0.1, 0.2, \ldots, 1)$ performs well without sacrificing computational efficiency. Third, to obtain more parsimonious measures that still can discriminate among different groups, we may select a subset of elements in the spectral envelope and optimal scalings that are most different among groups. This strategy has been used in \cite{fryzlewicz2009} for classifying nonstationary quantitative time series. For example, we compute 
\begin{equation*}\label{env}
\Delta(s) = \sum_{j=1}^{J} \sum_{h=j+1}^{J} \left[ \Lambda^{(j)}({\omega_s})  - \Lambda^{(h)}({\omega_s})   \right]^2,~~s=1,\ldots, K,
\end{equation*}
order $\Delta(s)$ decreasingly,  and then choose the top proportion of the elements in $\Delta(s)$. A leave-one-out cross validation approach that minimizes the classification error is then used to select an appropriate proportion.


Under Assumptions 1 and 4, classification consistency is established in Theorem  \ref{consist3}.
\begin{assumption}
	For fixed $m$ categories, $|| \Lambda^{(1)}-\Lambda^{(2)}||^2_2 + || \Gamma^{(1)}-\Gamma^{(2)}||^2_F \ge CT$ for a positive constant $C$.
\end{assumption}
\begin{theorem}\label{consist3}
	Under Assumptions 1 and 4, the probability of misclassifying $X_{r}$, a time series from group $\Pi_1$, to group $\Pi_2$, can be bounded as follows:
	$$
	P(D^{(r)}_{1, EnvSca} > D^{(r)}_{2, EnvSca}) = O(B_T^{2}T^{-2}),
	$$
	where $D^{(r)}_{1, EnvSca}$ and $D^{(r)}_{2, EnvSca}$ are defined in Equation \eqref{dist3}.
\end{theorem}		

\section{Simulation Studies}
\label{sec:sim}
We conduct simulation studies to evaluate performance of the proposed classification procedure. Following \cite{fokianos2003}, categorical time series $X_t$ are generated from the multinomial logit model as follows
\begin{equation*}
p_{t\ell} (\alpha) = \frac{\exp(\alpha_{\ell}' Y_{t-1})}{1+ \sum_{\ell=1}^{m-1}\exp(\alpha_{\ell}'Y_{t-1})},~~ \ell=1, \ldots, m-1,
\end{equation*}
and 
\begin{equation*}
p_{t m} (\alpha) = \frac{1}{1+ \sum_{\ell=1}^{m-1}\exp(\alpha_{\ell}'Y_{t-1})},
\end{equation*}
where $Y_{t}$ is a $(m-1)$-dimensional time series which has a one in the $\ell$th element if $X_t = c_{\ell}$ for $\ell=1,\ldots, m-1$ and zero elsewhere, $p_{t\ell}$  for $\ell= 1, \ldots, m$ are the probabilities of $X_t=c_{\ell}$ at time $t$ and satisfy $\sum_{\ell=1}^{m} p_{t\ell} = 1$, and $\alpha_{\ell}$ for $\ell= 1, \ldots, m$ are the regression parameters. The simulated model incorporates a lagged value of order one of $Y_t$ or $X_t$.  We consider three different cases under the multinomial model. For the first two cases, we let the number of categories $m=4$ and the number of groups $J=2$. For Case 1, we consider the following regression parameters.
\begin{eqnarray*}
\alpha_1 &=& (1.2, 1, 1)', \alpha_2 = (1, 1.2, 1)', \alpha_3 = (1, 1, 1.2)'  ~~ \text{if}~Y_t \in \Pi_1, \\
\alpha_1 &=& (0.3, 1, 1)', \alpha_2 = (1, 0.3, 1)', \alpha_3 = (1, 1, 0.3)'  ~~ \text{if}~Y_t \in \Pi_2.
\end{eqnarray*}
Figures \ref{simulated}(a) and \ref{simulated}(b) display realizations of time series from groups $\Pi_1$ and $\Pi_2$ in Case 1, respectively. For Case 2, the regression parameters are set to be 
\begin{eqnarray*}
\alpha_1 &=& (1.2, 1, 1)', \alpha_2 = (1, 0.8, 1)', \alpha_3 = (1, 1, 0.4)'  ~~ \text{if}~Y_t \in \Pi_1, \\
\alpha_1 &=& (0.4, 1, 1)', \alpha_2 = (1, 0.8, 1)', \alpha_3 = (1, 1, 1.2)'  ~~ \text{if}~Y_t \in \Pi_2.
\end{eqnarray*}
Figures~\ref{simulated}(c) and \ref{simulated}(d) present realizations of time series from groups $\Pi_1$ and $\Pi_2$ in Case 2, respectively. For Case 3, we consider $J=3$ different groups with the following regression parameters
\begin{eqnarray*}
\alpha_1 &=& (0.3, 1, 1)', \alpha_2 = (1, 0.3, 1)', \alpha_3 = (1, 1, 0.3)'  ~~ \text{if}~Y_t \in \Pi_1, \\
\alpha_1 &=& (1.2, 1, 1)', \alpha_2 = (1, 0.8, 1)', \alpha_3 = (1, 1, 0.4)'  ~~ \text{if}~Y_t \in \Pi_2, \\
\alpha_1 &=& (1.25, 0.5, 1)', \alpha_2 = (-2, -.75, -1)', \alpha_3 = (2, .75, -3)'  ~~ \text{if}~Y_t \in \Pi_3.
\end{eqnarray*}
100 replications are generated for the 27 combinations of 3 cases, 3 numbers of time series per group in the training data, $N_j =20, 50,100$ for all $j$, and 3 time series lengths $T=100,200,500$. A test dataset of 50 time series per group is generated for each repetition to evaluate the out-of-sample classification performance. Four different methods are implemented: the proposed classifier which utilizes both the spectral envelope and optimal scalings (EnvSca), the classifier using the spectral envelope only (ENV), the classifier using the optimal scalings only (SCA), and the sequence learner classifier (SEQ) of \cite{ifrim2011}.


Table \ref{tab:rate} summarizes the means and standard deviations of the correct classification rates. For Case 1, the proposed classifier and the envelope classifier perform similarly, and they both outperform sequence learner.  The scaling classifier has classification rates around 50\%, meaning that it is not better than a random guess. These results are unsurprising because $\Pi_1$ and $\Pi_2$ have different oscillatory patterns but similar traversals through categories, resulting in a poor classification rate if we use only the optimal scalings for classification. 
For Case 2, where the two groups are distinct mainly in the optimal scalings, the envelope classifier produces the lowest correct classification rate (around 50\%) among all methods considered. The proposed classifier and the scaling classifier perform similarly. They have slightly lower classification rates than sequence learner, which is designed to select and use all subsequences that are important in classifying responses and thus is well-suited for the setting in Case 2.
In Case 3, we consider three groups, and groups differ in cyclical patterns and scalings.  The proposed classifier has higher mean classification rates than the envelope and scaling classifiers. This is because groups are different in both oscillatory patterns and traversals through categories. The proposed classifier, by incorporating both the spectral envelope and optimal scalings, can produce better classification rates in this case.
It should be noted that sequence learner is developed under the framework of logistic regression and cannot classify a population of time series with more than two groups in its current form. One could extend sequence learner to multinomial logistic regression, but extensive programming efforts are needed and no prior results are available. Thus, we don't have simulation results for sequence learner in Case 3. 

In addition to classification, estimates of the tuning parameter $\kappa$ in the proposed algorithm allow for interpretable inference. For example, the average of estimated tuning parameters $\hat{\kappa}$ in our simulations for Cases 1, 2, and 3 are 1.00, 0.24, and 0.66, respectively.  This suggests that $\kappa$ can help us to identify whether groups are different in oscillatory patterns only, traversals through categories only, or a mixture of the two. 

\begin{table}[hbt!]
	{\begin{center}
			\begin{minipage}{14cm}
				\caption{\it \small Mean (standard deviation) of the percent of correctly classified time series across methods.}
				\centering
		\label{tab:rate}
				\vspace{0.5cm}
				\scalebox{.9}{
		\begin{tabular}{c|c|c|ccccc}
			Case & $N_J$ & T & EnvSca & SCA & ENV & SEQ \\[5pt]
			\hline
			& 
			
		 & 100 & 92.21 (3.41) & 49.42 (4.72)  & 93.32 (2.39) & 87.13 (3.32)  \\
			&20 & 200 & 96.91 (1.99) & 49.84 (4.60) & 98.16 (1.39)  & 93.24 (2.70) \\
			&~ & 500 & 98.78 (1.66) & 50.04 (4.71) & 99.98 (0.14) &  98.44 (1.40) \\
			\cline{2-7}
			& & 100 & 92.99 (2.68) & 49.92 (4.79) & 93.54 (2.28) & 90.40 (2.97) \\
			1 &50 & 200 & 97.64 (1.99) & 50.10 (4.31) &  98.47 (1.19) & 96.46 (2.04) \\
			&~ & 500 & 99.56 (0.64) & 49.63 (4.48) & 99.98 (0.14) & 99.56 (0.76) \\
				\cline{2-7}
			& & 100 & 93.68 (2.67) & 50.67 (5.00) & 93.76 (2.37) & 91.55 (2.71)\\
			&100 & 200 & 98.26 (1.30) & 49.73 (4.58) & 98.49 (1.19) & 96.73 (4.96)\\
			&~ & 500 & 99.80 (0.45) & 50.22 (4.72) & 99.97 (0.17) & 99.68 (0.60)\\
			\hline
			& 
			 & 100 & 71.13 (6.23) & 71.66 (6.00) & 50.42 (5.02) & 75.16 (4.45)\\
			&20 & 200 & 78.69 (5.76) & 79.30 (5.03) & 49.85 (5.29) & 83.32 (4.21) \\
			&~ & 500 & 88.27 (3.89) & 88.65 (3.96) & 49.94 (4.51) & 93.14 (2.58) \\
				\cline{2-7}
			& & 100 & 76.01 (5.36) & 76.25 (5.34) & 50.71 (4.39) & 77.94 (4.11)  \\
			2 &50 & 200 & 84.14 (4.03) & 84.22 (4.10) & 50.17 (4.92) &  86.71 (3.43)\\
			&~ & 500 & 94.20 (2.47) & 99.40 (2.34) & 50.93 (5.22) & 95.95 (2.23)\\
				\cline{2-7}
			& & 100 & 79.19 (4.60) & 79.48 (4.51) & 50.58 (4.83) & 78.56 (4.45) \\
			&100 & 200 & 87.59 (3.73) & 87.65 (3.67) & 39.61 (5.05) & 88.46 (3.32)\\
			&~ & 500 & 96.29 (1.83) & 96.31 (1.89) & 50.38 (5.04) & 96.68 (1.87)\\
			\hline
			& 
			 & 100 & 81.02 (4.69) & 70.43 (4.67) & 70.88 (3.97) & NA\\
			&20 & 200 & 89.64 (3.58) & 75.17 (3.48) & 80.61 (3.62) & NA\\
			&~ & 500 & 97.39 (1.80) & 81.80 (3.12) & 93.04 (2.27) & NA\\
				\cline{2-7}
			& & 100 & 83.79 (3.30) & 72.91 (3.67) & 71.08 (3.38) & NA\\
			3 &50 & 200 & 92.28 (2.62) & 78.18 (2.90) & 81.82 (2.91) & NA\\
			&~ & 500 & 98.42 (1.28) & 84.51 (3.02) & 94.32 (2.12) &  NA\\
				\cline{2-7}
			& & 100 & 84.97 (3.34) & 73.07 (3.29) & 71.37 (3.48)& NA\\
			&100 & 200 & 93.04 (2.09) & 79.99 (3.05) & 82.69 (2.87) & NA\\
			&~ & 500 & 98.67 (1.00) & 87.01 (2.59) & 94.29 (1.96) & NA
		\end{tabular}}
				\end{minipage}
			\end{center}}
		\end{table}	

\section{Analysis of Sleep Stage Time Series}
\label{sec:app}
During a full night of sleep, the body cycles through different sleep stages, including rapid eye movement (REM) sleep, in which dreaming typically occurs, and non-rapid eye movement (NREM) sleep, which consists of four stages representing light sleep (S1,S2) and deep sleep (S3,S4).  These sleep stages 
are associated with specific physiological behaviors that are essential to the rejuvenating properties of sleep.  Disruptions to typical cyclical behavior and changes in the amount of time spent in each sleep stage have been found to be associated with many sleep disorders \citep{zepelin05,altevogt2006sleep}.  Particular sleep disorders, such as nocturnal frontal lobe epilepsy (NFLE), are also difficult to accurately diagnose since clinical, behavioral, and electroencephalography (EEG) patterns for NFLE patients are often similar to those of patients with other sleep disorders, such as REM behavior disorder (RBD) \citep{mimic97, tinuper2017nocturnal}. Accordingly, there is a need for statistical procedures that can automatically identify cyclical patterns in sleep stage time series associated with specific sleep disorders and accurately classify patients with different sleep disorders.

The data for this analysis was collected through a study of various sleep-related disorders \citep{data} and is publicly available via \texttt{physionet} \citep{phy}. All participants were monitored during a full night of sleep and their sleep stages were annotated by experienced technicians every 20 seconds according to well-established sleep staging criteria  \citep{rk}.
We consider classifying sleep stage time series data collected from NFLE and RBD patients, for which differential diagnosis is particularly challenging \citep{tinuper2017nocturnal}. 
NFLE and RBD patients both experience significant sleep disruptions associated with complex, often bizarre motor behavior (e.g. violent movements of arms or legs, dystonic posturing) and vocalization (e.g. screaming, shouting, laughing), which is due to nocturnal seizures for NFLE patients \citep{tinuper2017nocturnal} and due to dream-enacting behavior in REM sleep for RBD patients \citep{rbddefn}. This makes differentiating RBD and NFLE patients particularly challenging. An objective, data-driven classification procedure that can automatically distinguish patients and aide differential diagnosis is needed.

The current analysis considers 8 hours of sleep stage time series from 
$N=46$ participants: 34 NFLE patients and 12 RBD patients.   This results in categorical time series of length $T=1440$ with $m=6$ sleep stages (REM, S1, S2, S3, S4, and Wake/Movement). Examples are provided in Figure~\ref{sleepstages}. In order to estimate the spectral envelope and optimal scalings, Wake/Movement is used as the reference category.
Leave-one-out (LOO) cross-validation is then used to empirically evaluate the effectiveness of the classification rule. For this data, the overall correct classification rate is 82.61\%, with 29 of the 34 NFLE patients correctly classified and 9 of the 12 RBD patients correctly classified.  The tuning parameter estimated via LOO cross-validation is $\hat{\kappa} = 0.852$.  This indicates that differences in spectral envelopes are relatively more important for accurately classifying members of each group compared to differences in optimal scalings for this data.

\begin{figure}[ht]
		\centering
		\includegraphics[scale=.56]{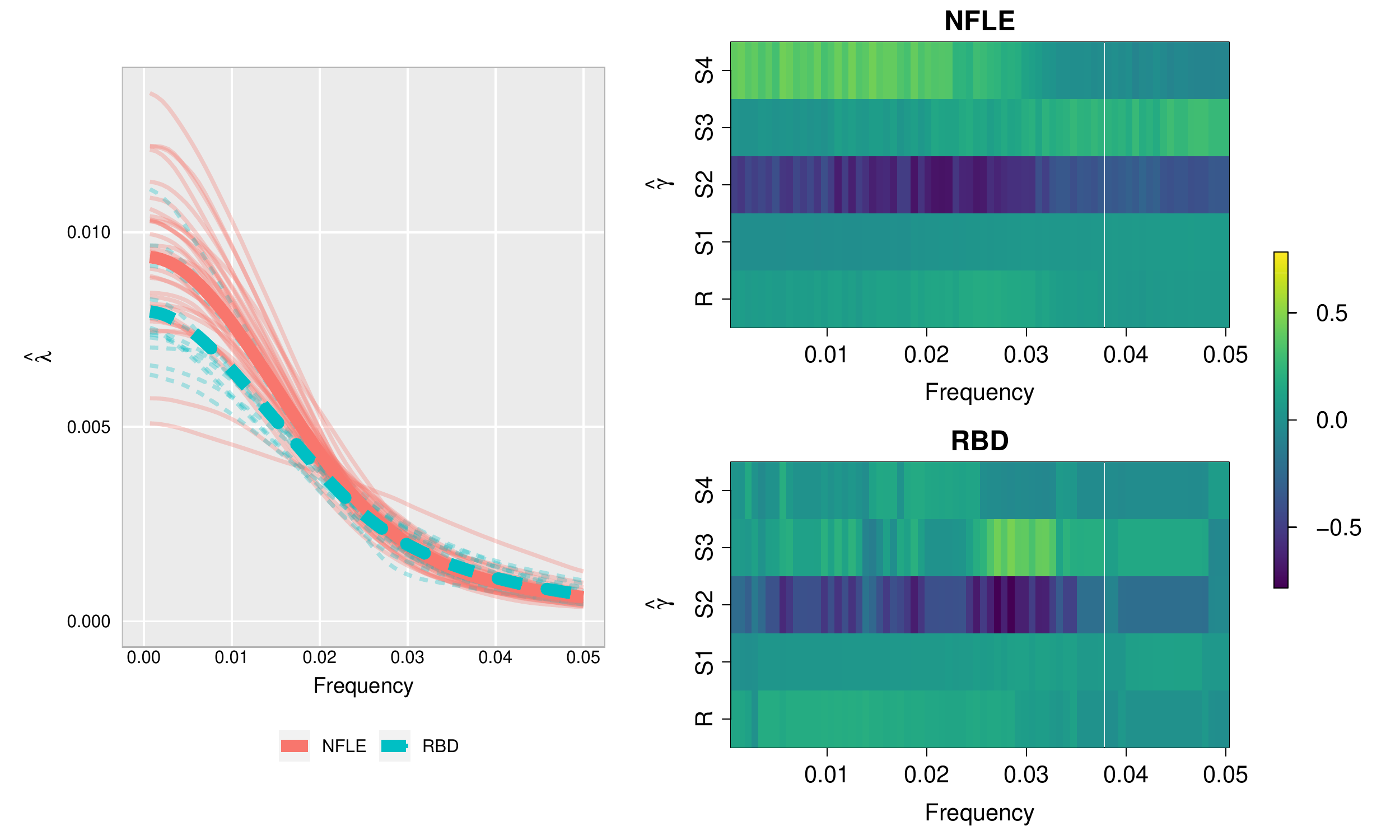}
		\caption{\small Left: Estimated spectral envelope for NFLE patients (solid red) and RBD patients (dashed blue) for low frequencies (below 0.05). Group-level estimated spectral envelopes are represented by the two thicker lines. Right: Estimated optimal scalings for NLFE patients (top) and RBD patients patients (bottom) for low frequencies (below 0.05).}
				\label{fig:sleep_class}
\end{figure}

In addition to providing a classification rule for categorical time series, the estimated group-level spectral envelopes and optimal scalings (see Figure \ref{fig:sleep_class})  provide insights into key differences in oscillatory patterns between the groups.  For both groups, power is concentrated at lower frequencies ($\le 0.05$) representing cycles lasting longer than $
6.7$ minutes and accounting for 87.4\% and 84.6\% of total power for the NFLE and RBD groups respectively.  This is expected as longer sleep cycles tend to dominate sleep, with typical NREM-REM sleep cycles lasting between 70 to 120 minutes \citep{altevogt2006sleep}.  Accordingly, our analysis focuses on differences between groups among low frequencies.  

First, the estimated spectral envelopes for the two groups (see Figure \ref{fig:sleep_class}) are reasonably well-separated for frequencies below 0.02 (representing cycles longer than 16.7 minutes), with NFLE patients generally exhibiting more low frequency power than RBD patients.  This result is not completely unexpected, since RBD patients tend to wake up abruptly at the end of a dream-enacting episode and are alert \citep{foldvary2013complex}, which can disrupt typical sleep cycles and reduce the prominence of low frequency oscillations.  On the other hand, NFLE patients do not typically wake up immediately following a nocturnal seizure \citep{foldvary2013complex}. 
The contrasting effects are also reflected in the data, in which RBD patients spend nearly twice as much time in the Wake/Movement stage during the night on average compared to NFLE patients (61.4 minutes vs. 32.1 minutes). 

\begin{figure}[ht]
		\centering
		\includegraphics[scale=.5]{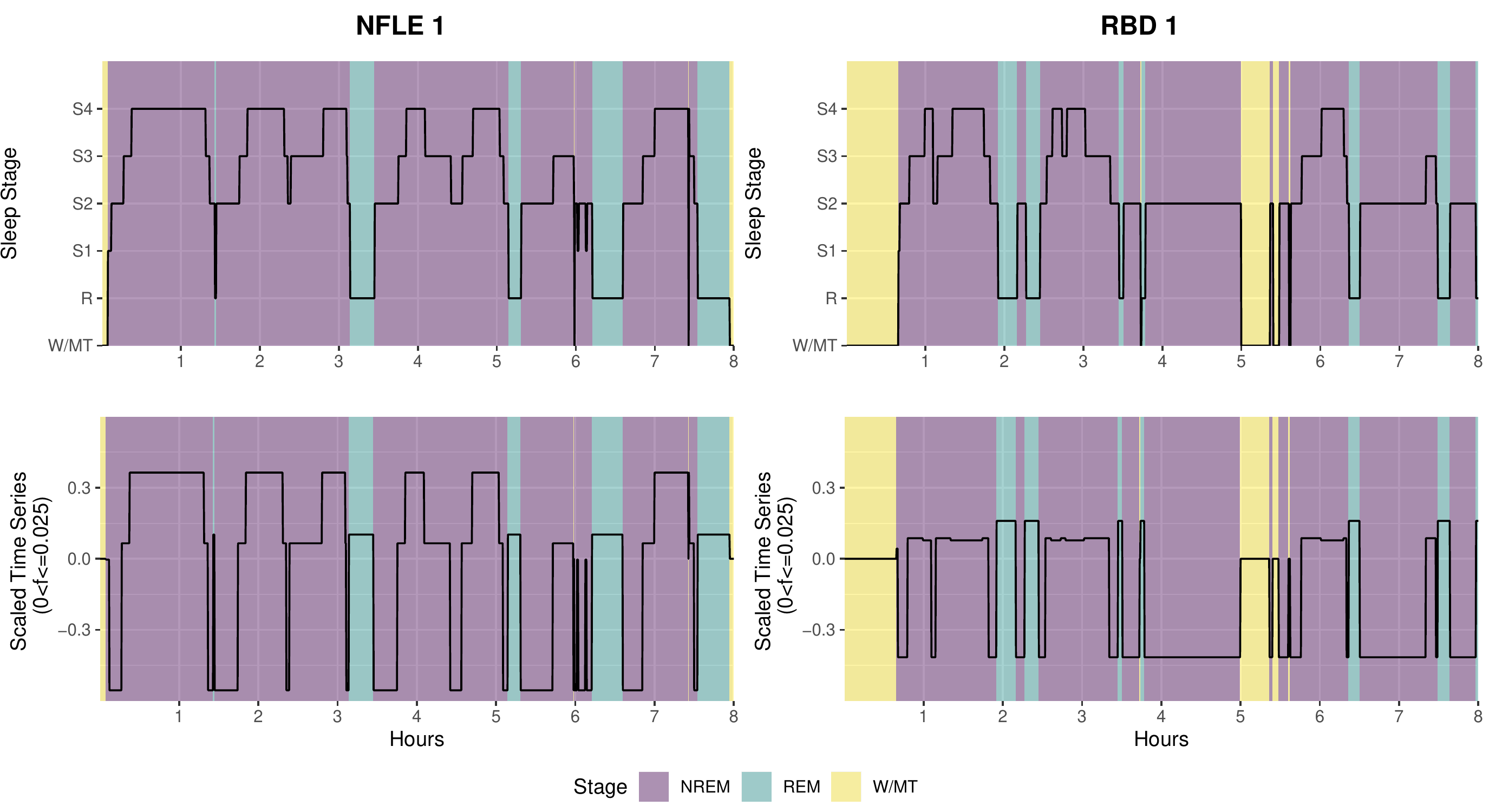}
		\caption{\small Top: Sample time series from the NFLE and RBD groups.  Bottom: Corresponding scaled time series based on the mean scaling for frequencies below 0.025 (i.e. cycles lasting more than 13 minutes).  Color corresponding to NREM (purple), REM (blue) and W/MT (yellow) sleep stages also provided.}
				\label{fig:scaltsapp}
\end{figure}

Second, differences in optimal scalings (see Figure \ref{fig:sleep_class}) are more subtle, with noticeable differences over some categories (e.g. S3, S4), but not all.  More specifically, scalings for frequencies below 0.025 indicate low frequency behavior in NFLE patients due to cycling among three broader sleep stage groupings: 1) light sleep (S2), 2) deep sleep (S4), and 3) a combination of transitional sleep stages (S1, S3), REM, and Wake/Movement.  On the other hand, RBD patients exhibit low frequency power primarily due to cycling in and out of light sleep (S2).  This can be attributed to more regular and prolonged periods of deep sleep (S4) observed in NFLE patients, lasting 14 minutes per onset and covering 20.9\% of total sleep on average, compared to RBD patients, lasting only 10.8 minutes per onset and covering 13.1\% of total sleep on average.  To better illustrate the differences in the optimal scalings, Figure \ref{fig:scaltsapp} provides a sample series from each group along with the scaled time series obtained by averaging optimal scalings over frequencies below 0.025.  Given the propensity for RBD patients to experience immediate sleep disruptions more so than NFLE patients, it is not surprising that RBD patients experience less deep sleep than NFLE patients.  

It is important to note that the proposed classification rule automatically adapts to these particular features of the spectral envelopes and optimal scalings through the data-driven estimate of $\hat{\kappa}=0.852$ using LOO cross-validation, which assigns more weight to differences in spectral envelopes in distinguishing between the two groups.  This is an important feature of the proposed classification procedure as it allows for the classification rule to adapt to differences between groups in the spectral envelope, optimal scalings, or both.

\section{Discussion}
\label{sec:end}
This article presents a novel approach to classifying categorical time series. An adaptive algorithm that utilizes both the spectral envelope and its corresponding set of optimal scalings for classification of categorical time series is developed. Classification consistency is also established. 
We conclude this article by discussing some limitations and related future extensions. First,  the proposed method assumes that the collection of time series is stationary. However, in some applications, the time series could be nonstationary, which would require time-varying extensions of the spectral envelope and optimal scalings for proper characterization. Incorporating nonstationarity may also further improve classification accuracy. A possible extension of the proposed method for classifying nonstationary categorical time series could use time-varying spectral envelope and scalings. Second, our method requires that all time series have the same length and all categories are observed.  However, in practice, time series may have different lengths and not all categories may be observed. For example, in the sleep study application, participants may have different lengths of full night sleep and some participants may not experience any movement during sleep.  Future research will focus on developing methods that can accommodate these kinds of time series observations. Third, our algorithm assumes that time series within the same group have the same cyclical patterns, while extra variability may be present in some applications \citep{krafty2016}. A topic of future research would be to incorporate within‐group variability into the classification framework.

\section*{Supplementary Material}
Supplementary material available online includes code for implementing the proposed classifier on the three cases of simulated data.

\section*{Appendix: Proofs}

To prove Theorem 1,2, and 3, we will make use of the following lemmas. 
\begin{lemma}
	Under Assumption 1 and assume that $h(\omega)^{re}$ has distinct eigenvalues. 
	Let $\lambda(\omega)$ and $\gamma(\omega)$ be the largest eigenvalue and corresponding eigenvector of $h(\omega)^{re}$. If $B_T \rightarrow \infty$ and $T \rightarrow \infty$ with $B_TT^{-1} \rightarrow 0$, then, 
	\begin{eqnarray*}
		|E \{ \hat{\lambda}(\omega) \} - \lambda(\omega)  | &=&  O(B_T T^{-1}), \\
		|E \{ \hat{\gamma}(\omega) \} - \gamma(\omega) |&=&  O(B_T T^{-1}).
	\end{eqnarray*}
\end{lemma}

\begin{lemma}
	Under Assumption 1 and assume that $h(\omega)^{re}$ has distinct eigenvalues. 
	Let $\lambda(\omega)$ and $\gamma(\omega)$ be the largest eigenvalue and corresponding eigenvector of $h(\omega)^{re}$. If $B_T \rightarrow \infty$ and $T \rightarrow \infty$ with $B_TT^{-1} \rightarrow 0$, then, 
	\begin{eqnarray*}
		|  \hat{\lambda}(\omega) - E \{  \hat{\lambda}(\omega) \} | &=&  O(B_T T^{-1}), \\
		| \hat{\gamma}(\omega) - E \{  \hat{\gamma}(\omega) \} |&=&  O(B_T T^{-1}).
	\end{eqnarray*}
\end{lemma}
Proofs of Lemma 1 and 2 are straightforward from \cite[Theorems 9.4.1 and 9.4.3]{brillinger2001} and thus omitted.


\subsection*{Proof of Theorem 1}
Recall that  $\hat{\lambda}= \{\hat{\lambda}(\omega_1), \ldots \hat{\lambda}(\omega_K)\}'$, where $K=\lfloor(T-1)/2\rfloor$, $D_{1, ENV} = || \hat{\lambda}-\Lambda^{(1)}||^2$, and $D_{2, ENV} = || \hat{\lambda}-\Lambda^{(2)}||^2$.  Let $\hat{\lambda}_s = \hat{\lambda}(\omega_s)$. It can be shown that 
\begin{equation*}
D_{1, ENV} - D_{2,ENV} = - 2 \sum_{s=1}^K (\hat{\lambda}_s - \lambda_s^{(1)}) (\lambda_s^{(1)} - \lambda_s^{(2)}) - \sum_{s=1}^{K}  (\lambda_s^{(1)} - \lambda_s^{(2)})^2.
\end{equation*}
It remains to show that 
\begin{equation}\label{detail}
P(D_{1,ENV} - D_{2,ENV} > 0) =  P\left ( \left [- 2 \sum_{s=1}^K (\hat{\lambda}_s - \lambda_s^{(1)}) (\lambda_s^{(1)} - \lambda_s^{(2)}) - \sum_{s=1}^{K}  (\lambda_s^{(1)} - \lambda_s^{(2)})^2
\right ] > 0 \right )
\end{equation}
is bounded. From Chebyshev inequality, we have
\begin{equation}\label{begin}
P(D_{1,ENV} - D_{2,ENV} > 0) \le \frac{E\left(\left [- 2 \sum_{s=1}^K (\hat{\lambda}_s - \lambda_s^{(1)}) (\lambda_s^{(1)} - \lambda_s^{(2)}) \right ]^2\right)}{ \left[\sum_{s=1}^{K} (\lambda_s^{(1)} - \lambda_s^{(2)})^2\right]^2}.
\end{equation}
Let's consider the numerator, 
\begin{eqnarray}\label{long}
&&E\left \{- 2 \sum_{s=1}^K (\hat{\lambda}_s - \lambda_s^{(1)}) (\lambda_s^{(1)} - \lambda_s^{(2)}) \right \}^2 = 4E\left \{ \sum_{s=1}^K (\hat{\lambda}_s - \lambda_s^{(1)}) (\lambda_s^{(1)} - \lambda_s^{(2)}) \right \}^2 \nonumber \\
&=&   4E\left \{ \sum_{s=1}^K (\hat{\lambda}_s - E(\hat{\lambda}_s) + E(\hat{\lambda}_s) - \lambda_s^{(1)}) (\lambda_s^{(1)} - \lambda_s^{(2)}) \right \}^2 \nonumber \\
&\le& 8E\left \{ \sum_{s=1}^K (\hat{\lambda}_s - E(\hat{\lambda}_s) ) (\lambda_s^{(1)} - \lambda_s^{(2)}) \right \}^2 + 8E\left \{ \sum_{s=1}^K ( E(\hat{\lambda}_s) - \lambda_s^{(1)}) (\lambda_s^{(1)} - \lambda_s^{(2)}) \right \}^2\nonumber \\
\end{eqnarray}	
Combine \eqref{begin} and \eqref{long}, we have $P(D_{1,ENV} - D_{2,ENV} > 0) \le I + II$, where 
\begin{eqnarray*}
	\RN{1} &=&\left. 8E\left \{ \sum_{s=1}^K (\hat{\lambda}_s - E(\hat{\lambda}_s) ) (\lambda_s^{(1)} - \lambda_s^{(2)}) \right \}^2 \middle/ \left[\sum_{s=1}^{K} (\lambda_s^{(1)} - \lambda_s^{(2)})^2\right]^2, \right. \\
	\RN{2} &=&\left. 8E\left \{ \sum_{s=1}^K ( E(\hat{\lambda}_s) - \lambda_s^{(1)}) (\lambda_s^{(1)} - \lambda_s^{(2)}) \right \}^2 \middle/ \left[\sum_{s=1}^{K} (\lambda_s^{(1)} - \lambda_s^{(2)})^2\right]^2, \right. \\
\end{eqnarray*}	
We analyze these two terms separately. For the first term \RN{1}, we have, the numerator $$8E\left \{ \sum_{s=1}^K (\hat{\lambda}_s - E(\hat{\lambda}_s) ) (\lambda_s^{(1)} - \lambda_s^{(2)}) \right \}^2 = O(B_T^2)$$ from Lemma 2. 
From Assumption 2, we have the denominator $\sum_{s=1}^{K} (\lambda_s^{(1)} - \lambda_s^{(2)})^2$ is of order $T^2$. Combine these results we have $\RN{1} = O(B_T^2 T^{-2})$. Similarly, using Lemma 1 and Assumption 2, we have $\RN{2} = O(B_T^2 T^{-2})$. Thus, complete the proof.

\subsection*{Proof of Theorem 2}
Recall that $\hat{\gamma}= \{\hat{\gamma}(\omega_1)', \ldots \hat{\gamma}(\omega_K)'\}'$, a $K \times (m-1)$ matrix, $D_{1, SCA} = || \hat{\gamma}-\Gamma^{(1)}||^2$ and $D_{2, SCA} = || \hat{\gamma}-\Gamma^{(2)}||^2$. It can be shown that
\begin{equation*}
D_{1, SCA} - D_{2,SCA} = - 2\sum_{\ell=1}^{m-1} \sum_{s=1}^K (\hat{\gamma}_{\ell,s}- \gamma_{\ell,s}^{(1)}) (\gamma_{\ell,s}^{(1)} - \gamma_{\ell,s}^{(2)}) - \sum_{\ell=1}^{m-1}\sum_{s=1}^{K}  (\gamma_{\ell,s}^{(1)} - \gamma_{\ell,s}^{(2)})^2.
\end{equation*}
we aim to show $P(D_{1, SCA} - D_{2,SCA} > 0)$. Similar to the proofs of Theorem I, we have 
$
P(D_{1, SCA} - D_{2,SCA} > 0) \le \RN{1} + \RN{2},
$
where 
\begin{eqnarray*}
	\RN{1} &=&\left. 8E\left \{ \sum_{\ell=1}^{m-1} \sum_{s=1}^K (\hat{\gamma}_{\ell,s} - E(\hat{\gamma}_{\ell,s}) ) (\gamma_{\ell,s}^{(1)} - \lambda_{\ell,s}^{(2)}) \right \}^2 \middle/ \left[\sum_{\ell=1}^{m-1} \sum_{s=1}^{K} (\gamma_{\ell,s}^{(1)} - \gamma_{\ell,s}^{(2)})^2\right]^2, \right. \\
	\RN{2} &=&\left. 8E\left \{ \sum_{\ell=1}^{m-1} \sum_{s=1}^K ( E(\hat{\gamma}_{\ell,s}) - \gamma_{\ell,s}^{(1)}) (\gamma_{\ell,s}^{(1)} - \gamma_{\ell,s}^{(2)}) \right \}^2 \middle/ \left[\sum_{\ell=1}^{m-1}  \sum_{s=1}^{K} (\gamma_{\ell,s}^{(1)} - \gamma_{\ell,s}^{(2)})^2\right]^2, \right. \\
\end{eqnarray*}	
Combine Lemma 1 and 2, and Assumption 3, we have $P(D_{1, SCA} - D_{2,SCA} > 0) = O(B_T^2 T^{-2})$.

\subsection*{Proof of Theorem 3}
We would like to show $P(D_{1,EnvSca} - D_{2,EnvSca} > 0) $ is bounded. It can be shown than 
$
D_{1, EnvSca} - D_{2,EnvSca} = A + B,
$
where 
$$
A =  \kappa \left[ \frac{-2 \sum_{s=1}^K (\hat{\lambda}_s - \lambda_s^{(1)}) (\lambda_s^{(1)} - \lambda_s^{(2)})}{\sum_{s=1}^K \hat{\lambda}_s^2 } - \frac{\sum_{s=1}^{K}  (\lambda_s^{(1)} - \lambda_s^{(2)})^2}{\sum_{s=1}^K \hat{\lambda}_s^2 } \right ],
$$
and 
$$
B = (1-\kappa)\left [\frac{-2\sum_{\ell=1}^{m-1} \sum_{s=1}^K (\hat{\gamma}_{\ell,s}- \gamma_{\ell,s}^{(1)}) (\gamma_{\ell,s}^{(1)} - \gamma_{\ell,s}^{(2)})}{\sum_{\ell=1}^{m-1} \sum_{s=1}^K \hat{\gamma}_{\ell,s}^2 } - \frac{\sum_{\ell=1}^{m-1}\sum_{s=1}^{K}  (\gamma_{\ell,s}^{(1)} - \gamma_{\ell,s}^{(2)})^2}{{\sum_{\ell=1}^{m-1} \sum_{s=1}^K \hat{\gamma}_{\ell,s}^2 }} \right ].
$$

Using the results in the proof of Theorems 1 and 2, and Assumption 4,  we have
\begin{eqnarray*}
&&P(A>0)\\ &&= P\left ( \left [- 2 \sum_{\ell=1}^{m-1}  \sum_{s=1}^K (\hat{\gamma}_{\ell,s} - \gamma_{\ell,s}^{(1)}) (\gamma_{\ell,s}^{(1)} - \gamma_{\ell,s}^{(2)}) - \sum_{\ell=1}^{m-1} \sum_{s=1}^{K}  (\gamma_{\ell,s}^{(1)} - \gamma_{\ell,s}^{(2)})^2
\right ] > 0 \right ) = O(B_T^2T^{-2}).
\end{eqnarray*}
and
\begin{eqnarray*}
P(B>0) = P\left ( \left [- 2 \sum_{s=1}^K (\hat{\lambda}_s - \lambda_s^{(1)}) (\lambda_s^{(1)} - \lambda_s^{(2)}) - \sum_{s=1}^{K}  (\lambda_s^{(1)} - \lambda_s^{(2)})^2
\right ] > 0 \right ) = O(B_T^2T^{-2}).
\end{eqnarray*}
Since 
\begin{eqnarray*}
P(D_{1, EnvSca} - D_{2,EnvSca}>0)  \le P(A>0) + P(B>0),
\end{eqnarray*}
we have the desired results. 
\bibliographystyle{asa}
\bibliography{locstatbib}
\end{document}